| | |
|---|---|
| Title: | Trapped Proton Environment in Medium-Earth Orbit (2000-2010) |
| Author(s): | Chen, Yue<br>Friedel, Reinhard Hans Walter<br>Kippen, Richard Marc |
| Intended for: | Report |
| Issued: | 2016-03-31 |

# Trapped Proton Environment in Medium-Earth Orbit (2000-2010)

Y. Chen, R.H.W. Friedel, Tom Cayton, and R. Marc Kippen[1]

*Los Alamos National Laboratory, Los Alamos, New Mexico, USA*

**Abstract:** This report describes the method used to derive fluxes of the trapped proton belt along the GPS orbit (i.e., a Medium-Earth Orbit) during 2000 – 2010, a period almost covering a solar cycle. This method utilizes a newly developed empirical proton radiation-belt model, with the model output scaled by GPS in-situ measurements, to generate proton fluxes that cover a wide range of energies (50keV- 6MeV) and keep temporal features as well. The new proton radiation-belt model is developed based upon CEPPAD proton measurements from the Polar mission (1996 – 2007). Comparing to the de-facto standard empirical model of AP8, this model is not only based upon a new data set representative of the proton belt during the same period covered by GPS, but can also provide statistical information of flux values such as worst cases and occurrence percentiles instead of solely the mean values. The comparison shows quite different results from the two models and suggests that the commonly accepted error factor of 2 on the AP8 flux output over-simplifies and thus underestimates variations of the proton belt. Output fluxes from this new model along the GPS orbit are further scaled by the ns41 in-situ data so as to reflect the dynamic nature of protons in the outer radiation belt at geomagnetically active times. Derived daily proton fluxes along the GPS ns41 orbit, whose data files are delivered along with this report, are depicted to illustrate the trapped proton environment in the Medium-Earth Orbit. Uncertainties on those daily proton fluxes from two sources are evaluated: One is from the new proton-belt model that has error factors < ~3; the other is from the in-situ measurements and the error factors could be ~ 5.

# Table of Contents

## 1. Introduction

The Medium-Earth-Orbit (MEO) region, the space loosely defined between the Low-Earth-Orbit region (LEO, with altitudes ~ 300 -1500 km) and the Geostationary orbit (GEO, with the altitude ~36000 km), is an ideal place to host global wireless and navigation satellite systems. Among them the GPS constellation is one of the most vital and precious space assets for the United States. Therefore, protecting the GPS fleets and other satellites in MEO from natural hazards creates an urgent need for capabilities of specification, understanding, and prediction of the ambient space environment in this region.

The GPS orbit, which is a typical MEO at an altitude ~20000 km, crosses the outer radiation belts 4 times daily and thus experiences a very hostile radioactive environment from both the trapped electron and proton belts, which are suspected to account for many anomalous radiation effects compared to other orbits. However, the MEO region has remained largely underexplored due to the past tendency of placing most satellites (and thus the measurement instruments) in LEO and GEO in order to avoid the intense region of radiation belts. Even with some measurements available, the data always have their own limitations, such as the limited energy coverage in the case of GPS satellites (see Section 4 for details), that cannot provide a full-spectrum description of the radiation belts. On the other hand, the de-facto standard empirical proton belt model AP8 [Sawyer and Vette, 1976] is based on age-old (1960s and '70s) data with limited coverage on L-shells and provides no description of the temporal behavior of fluxes other than long-term averages in the MEO region, which is contrary to observations showing large variations of protons in the outer belt modulated by geomagnetic storms [Gussenhoven et al., 1996]. Thus, the statistically averaged picture of the proton radiation belt from AP8 suffers from above model deficiencies.

This report aims to address the problem of characterizing the proton-belt environment in MEO region by providing long-term time series proton fluxes along one GPS satellite orbit. Since fluxes are given at a daily time scale, which is much longer than the average drift period of protons around the Earth (10s of minutes to several hours), we expect the results here to be very representative for the environment encountered in any GPS orbit. The method is described in Section 2. Section 3 presents the newly developed proton-belt empirical model, and the GPS ns41 proton measurements are introduced in Section 4. The report is concluded by showing results in Section 5. (For the electron-belt environment, see the other report by Tom Cayton).

## 2. Methodology

As mentioned in Section 1, neither in-situ measurements nor the existing statistical models can by themselves provide a reliable characterization of the dynamic trapped proton environment in the MEO region. To solve this problem, we apply the following 3-step method to derive dynamic trapped proton fluxes along the GPS orbit:

***Step 1:*** We develop a new proton radiation belt model based upon long-term measurements (1996-2007) from the CEEPAD experiment on board the Polar satellite. This model provides us an updated statistically averaged picture of proton flux distribution $\overline{J}(E,L,b)$, where *E* is proton energy, *L* is L-shell, and *b* is the ratio between magnetic field strengths at the spacecraft position and the magnetic equator along the same field line.

***Step 2:*** Using the flux measurement $j(E_0,L(t),b(t))_{ns41}$ at one energy channel $E_0$ along the GPS ns41 orbit at time *t*, we are able to derive the ratio between the measurement and model output flux at the satellite position

$$R(E_0,L(t),b(t))_{ns41} = \frac{j(E_0,L(t),b(t))_{ns41}}{\overline{J}(E_0,L(t),b(t))} \qquad (1)$$

where the ratio *R* varies by time and is a function of energy and satellite location.

***Step 3:*** Assuming the *R* from above Equation (1) is energy-independent (i.e., $R(L(t),b(t))_{ns41} = R(E_0,L(t),b(t))_{ns41}$), for any given energy point $E_i$ other than the energy channel $E_0$, we can derive proton fluxes by scaling the model output

$$J(E_i,L(t),b(t))_{ns41} = \overline{J}(E_i,L(t),b(t)) * R(L(t),b(t))_{ns41} \qquad (2)$$

along the GPS ns41 orbit during the period of 2000-2010 when ns41 measurements are available. Here Equation (2) implies that the spectral shape derived from the statistical model is valid, a necessary assumption due to the limited energy coverage of the GPS proton instrument (see Section 4), and thus only the amplitude is scaled.

Therefore, the derived fluxes $J(E_i,L(t),b(t))_{ns41}$ from above method keep the temporal feature from the GPS measurements, and also expand the energy range by using the energy spectra from the statistical model. In fact, we can further assume the ratio *R* is position-independent (i.e., $R(t) = R(E_0,L(t),b(t))_{ns41}$), therefore fluxes along other MEOs can also be obtained by using the radial distributions from the model. However, we constrain this report to fluxes along the GPS ns41 orbit only.

## 3. The New Trapped Proton Belt Model (PolarP)

Here we present a new empirical model for the proton radiation belt, which has been recently developed based upon long-term in-situ observations from the POLAR mission. This new **Polar P**roton empirical model (hereinafter called the "PolarP" model) aims to predict the range of trapped proton environment conditions that a satellite would encounter with a given orbit. This model provides proton flux values over a wide energy range (from 20 keV up to larger than 10 MeV) as a function of L-shell and equatorial pitch-angle. Comparing to the de-facto standard empirical model of AP8, this innovative model not only uses the latest available proton data, but also can provide statistical information of flux values such as occurrence percentiles instead of just mean values.

### 3.1. Instruments and Data Preparation

Long-term (1996-2007) proton observations used in the PolarP model are from the Comprehensive Energetic Particle and Pitch Angle Distribution (CEPPAD) experiment on board the Polar satellite. Polar has a highly elliptical (2 x 9 Re), 86° inclination orbit with a period of ~18 hr. The orbit plane precesses over all local times each year. During the mission, the apogee of Polar precesses from above the North Pole to above the South Pole, which makes the satellite (magnetic) equatorial crossing position varying from year to year (Figure 1). This is important since the pitch-angle coverage is obviously the best from magnetic equatorial crossing positions. Throughout its mission lifetime, Polar thus maps out all the geospace occupied by the GPS constellation.

Two subsystems of CEPPAD provide data used for PolarP development: The Imaging Proton Sensor (IPS) measures protons in 16 channels over the energy range of 20 - 1500 keV, with almost 4π solid angle coverage within a single 6 sec spin period; The High Sensitivity Telescope (HIST) has another 16 channels to measure protons over the energy range of 3.25 - 80 MeV. Data used here are pitch-angle resolved energy differential fluxes with 24 sec time resolution. To calculate magnetic coordinates, including the Roederer L-shell (i.e., the $L^*$ for a closed drift shell, [Roederer, 1970]) as well as the local and equatorial magnetic field strengths, we use IGRF as the internal magnetic field model and the Olson and Pfizer '77 model [Olson and Pfizer, 1977] for the external model. (The same magnetic field models are used through the report.)

The Polar instrument has its own set of limitations due to penetrating backgrounds (from cosmic rays, relativistic electrons, and the high energy trapped proton belt at low L-shells). In addition, the CEPPAD IPS instrument suffers from degrading and noisy detectors. Of the six detectors (with different look directions) all of them at varying time throughout the mission experience an increase in the noise threshold, making successively higher energy channels unusable. This limitation is alleviated by the fact that the 6 detectors provide redundant pitch angle information, and little coverage is lost if one or several detectors become unusable. However, towards the end of mission, as many as the lower 10 energies on most detectors have become unusable.

For Polar data preparation, we employed two strategies to combat these instrument limitations. For background, we used a shielded out-of-aperture detector in IPS to indicate highly penetrating fluxes. To be conservative, all data were discarded for any time that this background detector rose above 3 times the cosmic ray background level. As this occurs mostly in the high energy proton belt with L-shell below the GPS orbit, this is not s serious limitation. For detector noise, we developed an algorithm to detect the noise "floor" in each detector and energy. As long as there remained appreciable signal above this noise floor, we subtracted the noise floor from the data; Otherwise, when the noise got beyond this, the detector/energy combination was discarded and not used.

Besides the instrument limitations, several other limitations can also affect the reliability of the PolarP model. First, as shown in Figure 1, Polar measures different portion of the belts in each year, which makes it possible that the flux distributions are biased by dominance of data collected in one specific solar-cycle phase. How serious the bias can be depends on how much the solar cycle modulates the proton belt. In addition, since more major storms tender to occur during the inclining and declining phase of the solar cycle, modulation by magnetic storms can play a role here, too. Next, the quality of data from HIST channels (i.e., channels above MeV) is not as good as those from IPS, especially those for protons >20MeV. This makes the model output questionable for high energy end. Finally, since all the data come from one single instrument, this model heavily relies on the capabilities of CEPPAD. More will be discussed in Section 3.3.

## *3.2 PolarP Development and Model Results*

We first present the formalism of the model. The current version of the PolarP ignores any local-time dependence for simplicity. Thus, the physical space of the radiation belts is collapsed to 2 dimensions and is divided by many spatial bins with unequaled size. The center of each bin has the coordinates of ($L$, $\alpha$), where $L$ is the Roederer L-shell value for the drift shell and $\alpha$ is the equatorial pitch-angle. The equatorial pitch angle $\alpha$ is equivalent to the $b$ in Equations 1 and 2 since

$$\alpha = \arcsin(\sqrt{B_0/B}) = \arcsin(\sqrt{1/b}) \qquad (3)$$

where $B$ and $B_0$ are magnetic field strengths at local and equator along the field line. The equatorial pitch angle also determines the mirror latitude in the geomagnetic field, so any point of ($L$, $\alpha$) can also be transformed to the mirror point of (x, z) in the X-Z plane of a dipole magnetic field with the dipole pointing to the z-direction, as shown in Figure 2a. Contours of $L$ and $\alpha$ are also shown in Figure 2a and the values of $L$ and $\alpha$ grid points can be found in Table 1. For one energy range, the uni-directional flux values observed in one spatial bin are further constructed as an occurrence distribution of flux levels, as the example shown in Figure 2b. This flux level distribution enables one to calculate statistically the flux value at any percentile. For example, in Figure 2c, the flux value of ~$2\times10^8$/cm^2/s/sr/MeV at the 90$^{th}$ percentile indicates an observer has a 90% of chance of seeing flux levels at or below this value in this specified spatial bin and energy channel. In the case that the number of data points in one grid bin is below 100, all data in this bin are discarded so as to make the distributions statistically meaningful.

Examples of PolarP uni-directional flux distributions are presented in Figures 3 and 4. In Figure 3, panels in Column I show uni-directional flux distributions from the PolarP model at six energies (0.1, 0.5, 1.0, 1.5, 3.0, and 5.0 MeV, respectively) at the 50$^{th}$ percentile on the ($L$, $\alpha$) grids. As comparison, panels in Column II are flux distributions from AP8 Min model, and panels in Column III are flux ratios between AP8 and PolarP. Figure 4 shows uni-directional flux distributions at the 90$^{th}$ percentile for the PolarP model. Not surprisingly, the flux levels at the 90$^{th}$ percentile are higher than those at the 50$^{th}$. The AP8 distributions clearly show the one-belt structure, while the similar feature is also present in the PolarP distributions, indicating the contamination from energetic electrons in the outer radiation belt has been reduced to a reasonably low level. It should be noted that it is the Roederer L-shells calculated from the IGRF as well as Olson and Pfizer '77 (OP77) models that are fed to AP8 Min to obtain fluxes. This may introduce some errors (especially at regions with large radial gradients) since the L-shells in AP8

Min are calculated from the Jensen and Cain model [1962] and therefore can have slightly different values.

The PolarP model can also provide omni-directional fluxes, defined as the number of particles which arrive from all directions and traverse a test sphere with a unit cross-sectional area. Since uni-directional flux values along one $L$ contour provide the average flux equatorial-pitch-angle distribution, omni-directional fluxes at any point can be obtained by integrating uni-directional fluxes above the point along the same field line. Two such examples are shown in Figures 5 and 6, where again results from AP8 Min are shown for comparison.

From Figures 3 - 6, it can be clearly seen that the AP8 min flux values (i.e., the mean values) can be orders (up to ~2) of magnitude different from PolarP flux values not only at the $50^{th}$ percentile (i.e., the median values) but also the $90^{th}$ percentile. Specifically, in the MEO region (where AP8 Min and Max models are the same), for protons with energy lower than 1MeV, the AP8 flux values are obviously higher than those from PolarP; while for protons with energy higher than 1.5 MeV, with the limited coverage of L-shells from the AP8 model, the AP8 flux values tends to be lower than those from the PolarP. This suggests that the both the radial distributions and energy spectra from the AP8 are quite different from the PolarP. In addition, the comparison also indicates that the commonly accepted error factor of 2 on the AP8 fluxes over-simplifies and thus underestimates the variation of the proton belt. Thus, the AP8 model may be orders of magnitude off in estimating the time dose accumulation in the MEO regime.

### *3.3. Model Validation and Caveats*

As a first test, individual time series of Polar data are compared to the statistical results from the PolarP model. In Figure 7, Panels in Column I compare the time series of in-situ observations (black) during a ~ 5 hr period on one magnetic quiet day (1998/03/09) at four energies to the model flux values at seven percentiles $0^{th}$ $10^{th}$, $30^{th}$, $50^{th}$, $70^{th}$, $90^{th}$, and $100^{th}$ (gray-coded differently). It can be seen that all data points fall in the model flux ranges and most of the data are between the fluxes at the $10^{th}$ and $50^{th}$ percentiles. Panels in Column II repeat the comparison on a magnetically disturbed day (1998/03/11, the main phase of a magnetic storm with the minimum Dst ~ -120nT was in the previous day). It is shown that, during this disturbed period, proton fluxes at the two lowest energies move up to PolarP model fluxes at higher percentiles, while not much change to the high energy protons. This reflects the dynamic behavior of the proton belt and suggests the possibility of new protons being injected into the belt during the storm (another possibility is the adiabatic effects causes the change). Dates in Figure 7 are randomly picked and repetition of above has shown similar results on other days (not shown here). This validation step demonstrates that there is no significant algorithm or coding errors in the model development steps.

Further validation involves the comparison of PolarP fluxes to the independent GPS ns41 data set and one example is shown in Figure 8. From this figure, it can be seen that ns41 fluxes are lower than PolarP fluxes at the $10^{th}$ percentile during the whole day 345 of year 2000 (a geomagnetically quiet day), which is not quite the same as shown in Figure 8. However, the inter-calibration between Polar CEPPAD and ns41 BDD-IIR indicates that CEPPAD fluxes are systemically ~ 2 - 10 times higher than ns41 BDD-IIR measurements (see Section 4 for details). Therefore, by applying a multiplying factor of 5 on GPS data (or, alternatively, dividing PolarP fluxes by a factor of 5), we would expect to see adjusted GPS data falling between PolarP fluxes at $30^{th}$ and $50^{th}$ percentiles during at its equatorial crossings, as is consistent with the comparison with CEPPAD data in above step.

There are caveats that one should be aware of when using the PolarP model. As mentioned in Section 3.1, data used in this model have their own limitations, which are passed on to this model. For example, the background and contamination of energetic electrons (especially in the outer belt) on the instrument may not be removed entirely. In addition, since the statistical distributions come from measurements from one single instrument, this model heavily relies on the capabilities of CEPPAD. For instance, the maximum flux level on the $100^{th}$ percentile for one energy channel may only reflect the saturated flux level measurable by the channel. The real maximum flux level can be even higher. Since the model development is constructed in an open way, it will be easy to incorporate improved CEPPAD data with high quality and other data sets so as to keep on improving the PolarP model in the future.

### *3.4 Summary of the PolarP Model*

This concludes the Step 1 as described in Section 2. We present the new empirical proton-belt model, the PolarP model, which is based upon long-term in-situ observations from the Polar CEPPAD. Since this model provides proton flux values over wide ranges of energies (from 20 keV up to larger than 10 MeV), L-shells (up to ~10), and equatorial pitch-angles, the PolarP is able to provide the trapped proton environment conditions that a space environment instrument would encounter with most given orbits. Comparing to the de-facto standard empirical model of AP8, this innovative model not only uses the latest available data , but also can has new features such as providing statistical distributions of fluxes instead of just mean values. The comparison also shows quite different flux results from the two models and indicates that the commonly accepted error factor of 2 on the AP8 flux output over-simplifies and thus underestimates the variation of the proton belt.

## 4. GPS ns41 Proton Measurements

Each GPS satellite has a circular orbit with a radius of 4.2 Re, inclination of 55 degree, and 12 hr period. Thus the satellite travels across the radiation belts every ~ 6 hrs (with ~3 hrs inside the belts) at L-shells ≥ ~ 4 at different local time.

The Burst Detector Dosimeter (Block) IIR (BDD-IIR) [Cayton *et al.*, 1998] onboard GPS ns41 provides eight energy channels to measure proton fluxes from 1.3 MeV up to > 54 MeV. The sensor collimators have an $110^{o}$ field of view. Since the satellite is three-axis stabilized and always points to the Earth, the measurement of equatorial pitch angle range varies along the orbit and is closest to the 90$^o$ at equatorial crossings. Here in this work we use 4 min pitch-angle averaged proton fluxes from the lowest energy channel (1. 27 – 5.3 MeV). The reason we do not use data from other channels is because the counts for higher energies are seldom higher than the background and thus are statistically insignificant and unreliable.

Figure 9 provides a close look of GPS ns41 proton data during a 5-day interval. Panel A plots all proton fluxes (black curve) in time series and points with L-shell smaller than 6 are highlighted by diamond symbols in gray. Panel B shows the L-shells sampled by the spacecraft and Dst and Kp indices in Panel C indicate this 5-day interval being magnetically quiet. It can be clearly seen that, during each GPS crossing of the radiation belt (4 times a day), the proton flux has the highest value at the equator ($L^* \sim 4$) and then reduces quickly when moving to high L-shells. It is the consequence of several factors. First, the gradient in radial distributions and the gradient in pitch-angle distributions determine that the proton flux gets lower at higher altitude (i.e., larger L-shell). In addition, since BDD-IIR does not cover the whole $4\pi$ solid angle space and the satellite always points to the Earth, protons sampled by the GPS are further away from the 90$^o$ local pitch angle at high L-shells, which enhances the decrease of flux values off equator since trapped proton flux is highly peaked around the 90$^o$ local pitch angle. Together those factors account for more than three orders of magnitude variation of proton fluxes observed by GPS in each radiation belt crossing.

Figure 10 provides an overview of ns41 proton data during the whole period from 2000 to present (2010). There are two noticeable features by looking at variations of fluxes measured at equators (i.e., the top envelop line of fluxes in Panel A). First, many variations are associated with magnetic storms (indicated by decreasing Dst index in Panel B): Proton fluxes tend to drop during storm main phases and then recover afterwards. The other is that the average proto flux level in the second half of this period is ~ 5 times higher than that in the first half. This is suspected to be a consequence of the solar minimum period – fewer storms lead to less de-trapping of protons which in turn can build up to higher levels. Both features remind us that a static proton model is definitely not enough to describe proton dynamics in the MEO region.

Preliminary inter-instrument calibration between CEEPAD and BDD-IIR proton channels has been conducted by comparing measurements from both instruments during solar proton events (SPEs) in the open-field-line region (defined as the place where no closed drift shell can be found by tracing magnetic field lines in the global magnetic field models and thus no stably trapped protons). During SPEs, highly energetic protons ejected from the Sun sweep across the Earth magnetosphere and occupy the open-field-region, in which each particle instrument should measure the same flux value no matter its location. Figure 11 depicts the fluxes measured simultaneously by both CEEPAD and BDD-IIR with proton energy between [1.27, 5.3] MeV in SPEs during 2001 – 2005, when a list of SPEs is available with ~ 60 events in total). It is obvious that CEPPAD measures fluxes ~ 2 – 10 times higher than BDD-IIR does during those events. It is admitted that this inter-instrument calibration is not final since factors, such as solar proton energy spectrum, instrument response, and others, are not considered here. But at lease it reveals that a non-trivial measurement difference between the two instruments does exist and Figure 12 further confirms this difference.

In Figure 12, Panel C shows the averaged ratios $R$ (as defined by Equation (2)) between BDD-IIR fluxes and the corresponding PolarP fluxes at the 50$^{th}$ percentile at GPS equatorial crossings during the whole period. The average is performed over a running 27-day window. The ratios indicate that the PolarP fluxes are systemically higher than ns41 BDD-IIR measurements. During 2001 – 2005, i.e., DOY ~ 0-1825 in the panel, the majority of averaged ratios falls between 0.1 and 0.5 and has a mean ratio ~ 4.4, consistent with results shown in Figure 11. (Considering the PolarP is a statistic model, the increment of ratios in the rest of the period is due to reasons other than inter-calibration, most likely the enhancement of trapped proton belt in MEO.)

The difference between the two instruments thus puts an uncertainty on in-situ measurements from either CEPPAD or BDD-IIR. This uncertainty in in-situ measurements is counted as a potential error source in this report. However, we emphasis that so far we still cannot tell the measurement from which instrument is closer to the "real" fluxes without further investigation. We should mention that all the BDD-IIR calibrations have been done in house to a high degree of fidelity and fluxes from GPS ns41 agree with AP8 model quite well (not shown here). Additionally, GPS ns41 data are not replaceable in this work since they are the only available long-term in-situ measurements in MEO region. In contrast, nominal flux conversion factors for CEPPAD are used for this work and we currently do not have the resources to perform a comprehensive analysis of the CEPPAD responses to backgrounds so that residual contamination on CEPPAD data may still exist. Consequently, in this work we put our trust on GPS data and will use these data to scale the PolarP results to match the GPS measurements.

## 5. Derivation of Extended Proton Fluxes along GPS ns41 Orbit

With both PolarP model and ns41 data in hand, we proceed to the Step 2 as described in Section 2. Following Equation (1), the ratios $R$ are derived between ns41 fluxes and PolarP model omni-directional fluxes, and the latter are obtained by feeding ns41 Roederer L-shell and $B/B_0$ values calculated from OP77 model to the PolarP model. Figure 12 plots the ratios as a function of time. Panel A shows ratios during the whole period, and Panel B shows details by plotting ratios in a 10-day interval. Here we use PolarP omni-directional fluxes at the 50$^{th}$ percentile. In addition, the same ratio is used for each (or two consecutive) radiation belt crossing(s) as shown in Panel B and the ratio value is obtained from fluxes when ns41 travels across the magnetic equator with $L^* \sim 4$. This simplification can be justified by three reasons: One is that, as mentioned in Section 4, GPS samples fluxes off 90$^o$ local pitch angle at off-equator positions and thus not measure the core-part of the comparable omni-directional fluxes; Next, modulation of the proton belt by magnetic storms usually has a time scale larger than 6 hrs; Last, the final product needed is daily integrated flux which justifies the average over 6 hrs. Panel C plots the averaged $R$ as discussed in Section 4.

It should be noted that we filter out the ratios during transient SPEs since we focus on the trapped proton environment in this work. It is done by picking out dates when ns41 flux levels are unusually high in the open-field-line region during SPEs. Original ratios between GPS and PoalrP model fluxes during those dates are discarded and replaced by ratios available in the previous latest non-SPE dates. Trapped SPE protons are kept as long as the proton flux level in the open-field-line region goes back to the normal low level.

Finally, with the PolarP model and the ratios calculated above, by using Equation (2), we derive the trapped proton fluxes (i.e., the delivered data sets) for 16 energies along the ns41 orbit from 2000 to present (2010), as shown in Figure 13. In this way, the derived fluxes keep the temporal feature from the ns41 measurements, and expand the energy coverage by using the energy spectra from the PolarP model. It should be noted that the fluxes in Figure 13 are based on PolarP fluxes at the $50^{th}$ percentile. Derived fluxes values can be different if using PolarP fluxes at other percentiles. Figure 14 presents one example, which shows the ratios between derived fluxes based upon PolarP fluxes at the $90^{th}$ and $50^{th}$ percentiles. The non-unity ratios are caused by the different energy spectra at the two percentiles. Obviously, PolarP fluxes at the $90^{th}$ percentile have a harder energy spectrum than that at the $50^{th}$, since the ratios are larger than 1 for MeV protons and lower than 1 for 10s and 100s keV protons. Consequently, the flux ratios shown in Figure 14 can be used as the error factors from the PolarP model for the delivered proton fluxes.

One application of the derived extended proton fluxes is to calculate the accumulative fluence in MEO orbits. Three such examples are shown in Figure 15, which plots the fluence accumulated over the whole period (2000 – 2010) along the ns41 orbit for low- ($\geq 0.1$ MeV), medium- ($\geq 1$MeV) and high-energy ($\geq 5$MeV) trapped protons, respectively. As comparison, fluence curves calculated based on the AP8 Min model are also presented: The dotted lines show AP8 results by using Roederer L-shell values calculated from the OP77 model as the input parameter (the same L-shell definition used in the PolarP model); the dashed lines show AP8 results by using McIlwain L-shell values as the input parameter. It can be clearly seen that each AP8 curve has a constant slope, i.e., a constant fluence accumulative speed, since AP8 is a static model, while the slope of the fluence curve based on derived proton fluxes changes since the flux level varies day by day as shown in Figure 13. For instance, in Panel A, the high extended flux levels in the second half of the period makes the fluence catch up and eventually exceed that from AP8. Another feature shown in Figure 15 is the energy-dependent difference between the fluence curves based upon AP8 and extended proton fluxes. For example, in the beginning of the period, the AP8 overestimates fluence for low and medium energy protons (Panels A and B), but underestimates fluence from high-energy protons (Panel C). This reflects the different energy spectra from the AP8 model and PolarP model, which is consistent with our previous discussion in Section 3.2. It should be noted that the AP8 model is constructed using the McIlwain L-shells from the Jensen and Cain model [1962] (which we do not have in the house unfortunately) so that none of the AP8 curves shown here are precisely right. In fact, as mentioned in Section 3.2, using different definitions of L-shell calculated from different magnetic field values can lead to different values in L-shells, and thus introduce difference in the AP8 output flux (and fluence) values, as shown in Figure 15. The difference can be non-trivial at places with large model flux radial gradients, as the medium-energy protons in MEO shown by Panel B. However, although it is not exactly an orange-to-orange comparison in Figure 15, the difference between the AP8 model and extended proton fluxes does exist and is significant.

In summary, this report describes a 3-step method used to derive fluxes of the trapped proton belt along the GPS ns41 orbit during a solar-cycle-long period (2000 - 2010). This method utilizes a newly developed empirical proton radiation-belt model, with the model output scaled by GPS ns41 in-situ measurements, to generate proton fluxes that cover a wide range of energies (50keV- 6MeV) and bear temporal features as well. Derived daily proton fluxes along the ns41 orbit, whose data files are delivered along with this report, are depicted to illustrate the trapped proton environment on a Medium-Earth Orbit. Uncertainties on those daily proton fluxes come from two sources: One is from the PolarP model and preliminary results show error factors < ~3; the other is from the in-situ measurements and the error factor could be ~5 but further investigation is needed.

**Acknowledgements** This work was supported by the MEORAD project funded by USAF/SMC and Project 1131871 funded by NSF National Space Weather Program.

# References

Cayton T. E., D.M. Drake, K.M. Spencer, M. Herrin, T.J. Wehenr, and R.C. Reedy, Description of the BDD-IIR: Electron and proton sensors on the GPS. *Technical Report* **LA-UR-98-1162**, Los Alamos National Laboratory, Los Alamos, NM, 87545, USA, 1998

Gussenhoven, M.S., E.G. Mullen, and D.H. Brautigam, Improved understanding of the Earth's radiation Belts from the CRRES satellite", *IEEE Trans. Nucl. Sci.*, **43**, 353-368, 1996

Jensen, D.C. and J.C. Cain, An interim geomagnetic field, *J. Geophy. Res.*, **67**, 3568, 1962

Olson, W., and K. Pfizer, Magnetospheric magnetic field modeling, Tech. Rep., McDonnell Douglas Astronaut. Co, Huntington Beach, Calif., 1977

Roederer, J.G., Dynamics of geomagnetically trapped radiation, Springer, Verlag, New York, 1970

Sawyer, D. and J. Vette, AP-8 trapped proton environment for solar maximum and solar minimum, National Space Science Data Center, **Report 76-06**, Greenbelt, Maryland, 1976.

Vette, J.I., The AE-8 Trapped electron model environment, *Tech. Rep.* **NSSDC/WDC-A-R&S91-24**, 1991

**Table 1: Summary of Grid Points in the PolarP Model**

| L-shell Grid Points (Total number 23): | | | | | | | |
|---|---|---|---|---|---|---|---|
| 1.10 | 1.25 | 1.35 | 1.45 | 1.60 | 1.80 | 2.00 | 2.40 |
| 2.80 | 3.20 | 3.60 | 4.00 | 4.40 | 4.80 | 5.25 | 5.75 |
| 6.25 | 6.75 | 7.50 | 9.00 | 11.0 | 13.0 | 14.0 | |
| | | | | | | | |
| Equatorial Pitch Angle Grid Points (In degree, total number 26): | | | | | | | |
| 90.00 | 82.94 | 76.18 | 69.70 | 63.50 | 57.60 | 51.98 | 46.66 |
| 41.62 | 36.86 | 32.40 | 28.22 | 24.34 | 20.74 | 17.42 | 14.40 |
| 11.66 | 9.216 | 7.056 | 5.184 | 3.600 | 2.304 | 1.296 | 0.5760 |
| 0.1440 | 0.0000 | | | | | | |
| | | | | | | | |
| Energy Points (In MeV, total number 30) | | | | | | | |
| 1.89e-2 | 2.44e-2 | 3.24e-2 | 4.31e-2 | 5.72e-2 | 7.61e-2 | 1.02e-1 | 1.38e-1 |
| 1.89e-1 | 2.59e-1 | 3.55e-1 | 4.89e-1 | 6.74e-1 | 9.29e-1 | 1.28 | 1.73 |
| 4.17 | 5.19 | 6.46 | 8.04 | 1.00e1 | 1.25e1 | 1.55e1 | 1.93e1 |
| 2.41e1 | 3.00e1 | 3.73e1 | 4.64e1 | 5.78e1 | 7.20e1 | | |
| | | | | | | | |
| Flux units: #/cm$^2$/s/sr/MeV for uni-directional fluxes and #/cm$^2$/s/MeV for omni-directional fluxes | | | | | | | |

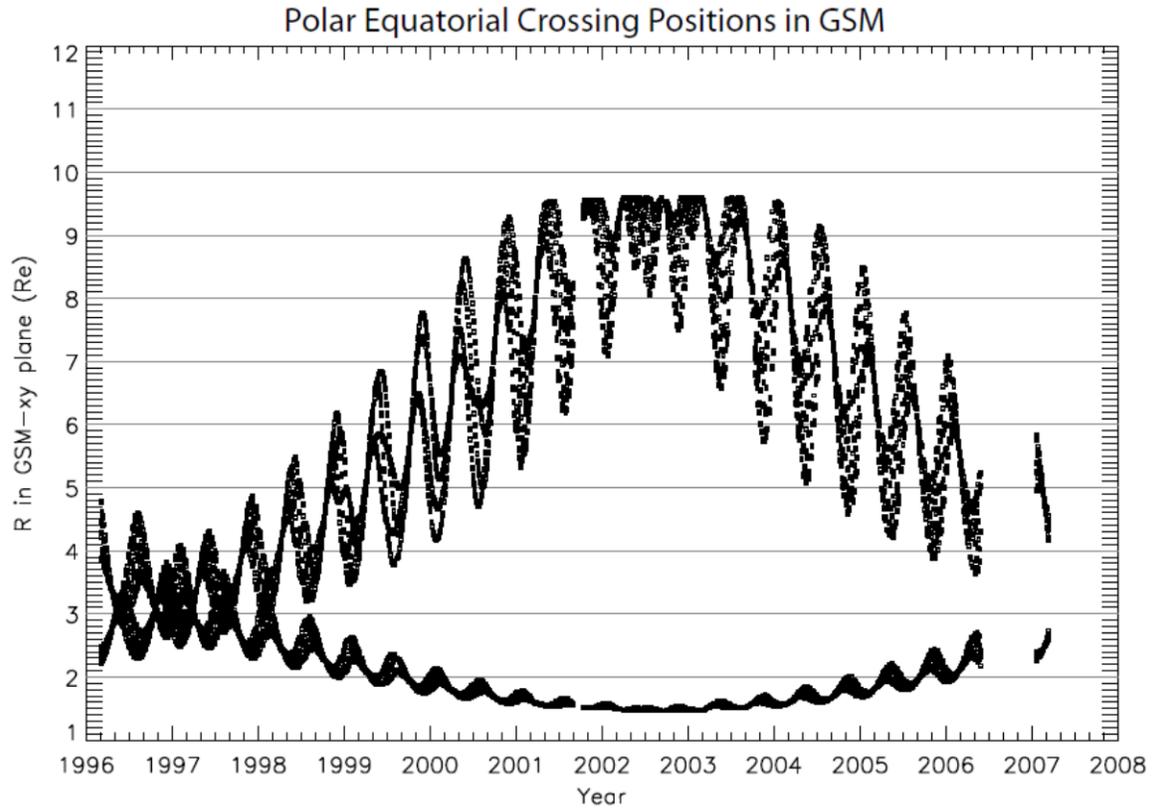

**Figure 1 Polar equatorial crossing positions in GSM coordinates during the mission.** Here the equatorial crossing is defined as the spacecraft GSM z-component value being < 0.2 Re, which is different from but can approximate the magnetic equatorial crossing.

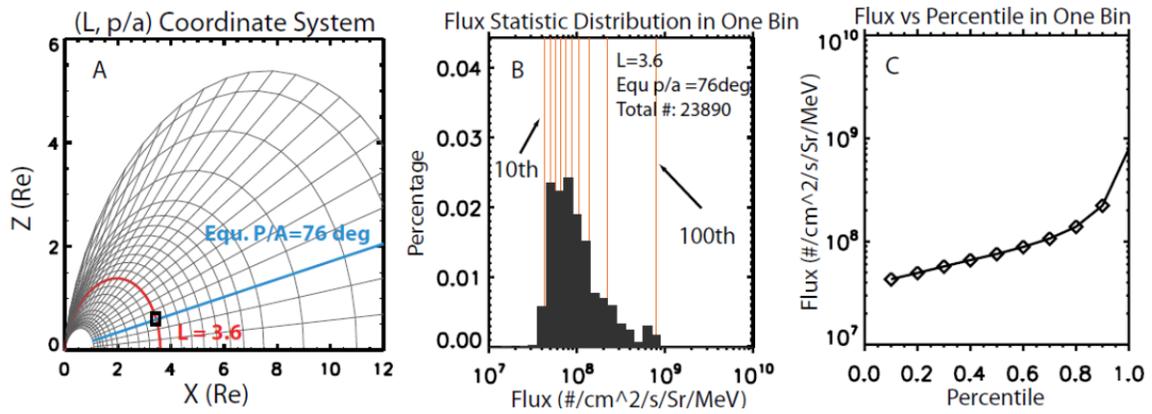

**Figure 2 Formalism of the PolarP model. A)** Contours of *L* and equatorial pitch-angle *α* in the X-Z plane with a magnetic dipole in z-direction. The curve for L = 3.6 is highlighted in red as one example and the curve for α = 76° is in blue. The black square indicates a spatial bin whose contents are shown Panel B and C. **B)** The flux statistical distribution in one bin (occurrence percentages normalized to the total number of flux data points in this bin). The vertical red curves indicate the flux levels for 10 percentiles (accumulative occurrence percentages) from the 10$^{th}$ to the 100$^{th}$ with an increment of 10. **C)** Flux levels vs. percentiles in the same bin.

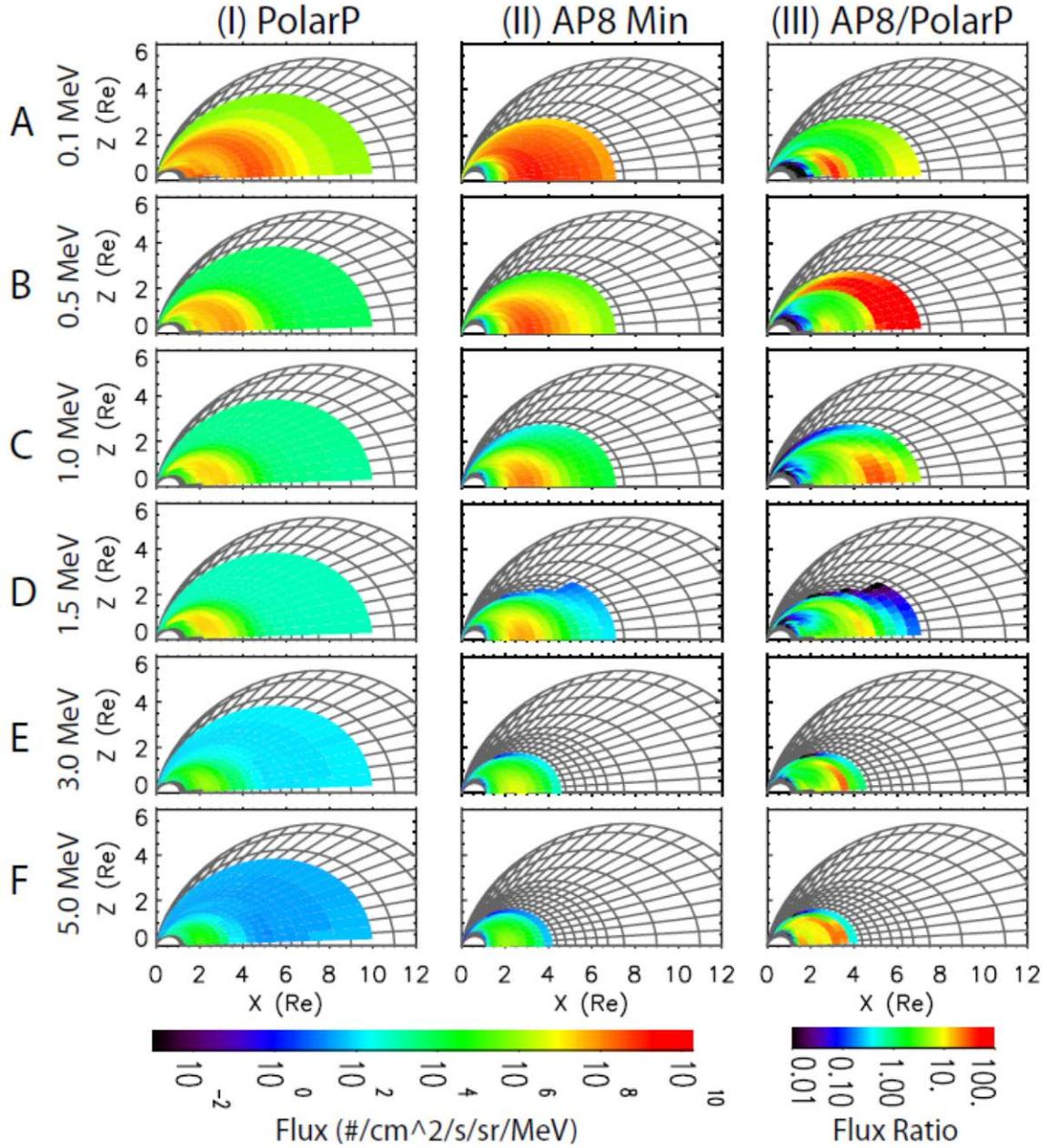

**Figure 3 Uni-directional flux distributions from the PolarP model at the 50th percentile and the AP8 Min model. Column I)** From top to bottom, PolarP uni-directional flux distributions at the 50th percentile at six energies, 0.1, 0.5, 1.0, 1.5, 3.0, and 5.0 MeV, respectively. **Column II)** AP8 Min uni-directional flux distributions at the same six energies. **Column III)** Ratios between AP8 Min and PolarP.

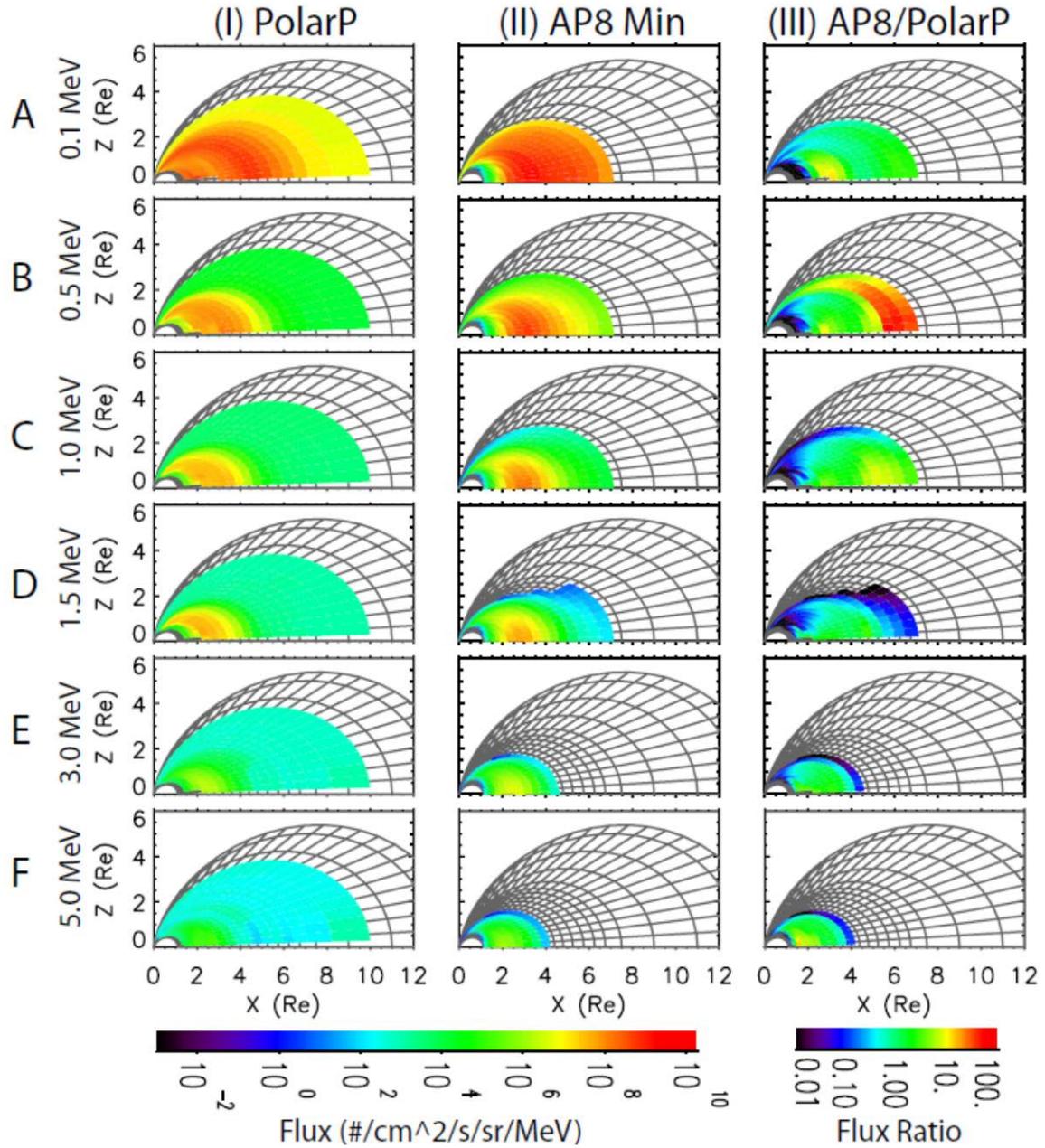

**Figure 4 Uni-directional flux distributions from the PolarP model at the 90th percentile and the AP8 Min model. Column I)** From top to bottom, PolarP uni-directional flux distributions at the 90th percentile at six energies, 0.1, 0.5, 1.0, 1.5, 3.0, and 5.0 MeV, respectively. **Column II)** AP8 Min uni-directional flux distributions at the same six energies. **Column III)** Ratios between AP8 Min and PolarP.

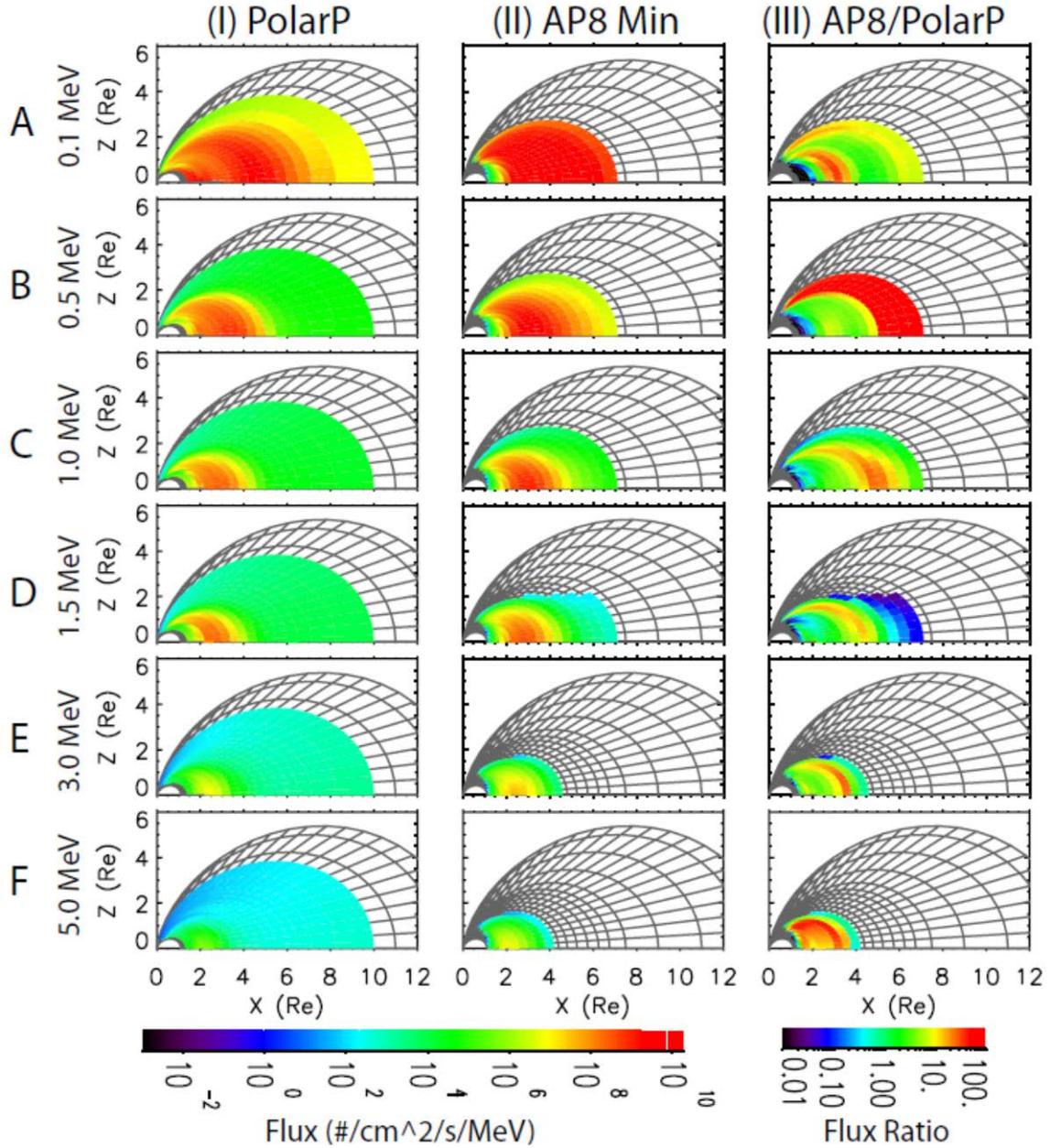

**Figure 5 Omni-directional flux distributions from the PolarP model at the 50th percentile and the AP8 Min model. Column I)** From top to bottom, PolarP omni-directional flux distributions at the 50th percentile at six energies, 0.1, 0.5, 1.0, 1.5, 3.0, and 5.0 MeV, respectively. **Column II)** AP8 Min omni-directional flux distributions at the same six energies. **Column III)** Ratios between AP8 Min and PolarP.

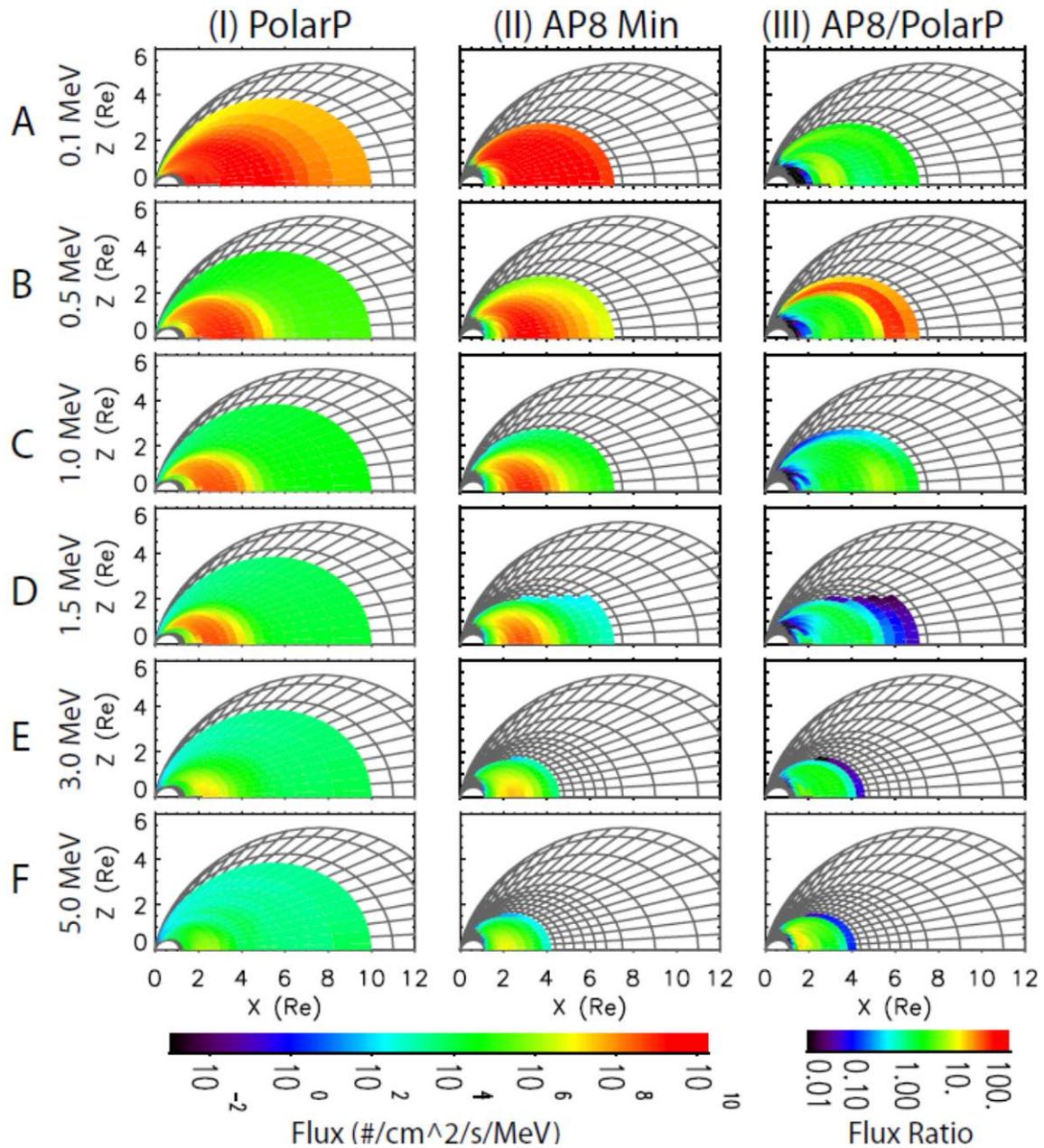

**Figure 6 Omni-directional flux distributions from the PolarP model at the 90$^{th}$ percentile and the AP8 Min model. Column I)** From top to bottom, PolarP omni-directional flux distributions at the 90$^{th}$ percentile at six energies, 0.1, 0.5, 1.0, 1.5, 3.0, and 5.0 MeV, respectively. **Column II)** AP8 Min omni-directional flux distributions at the same six energies. **Column III)** Ratios between AP8 Min and PolarP.

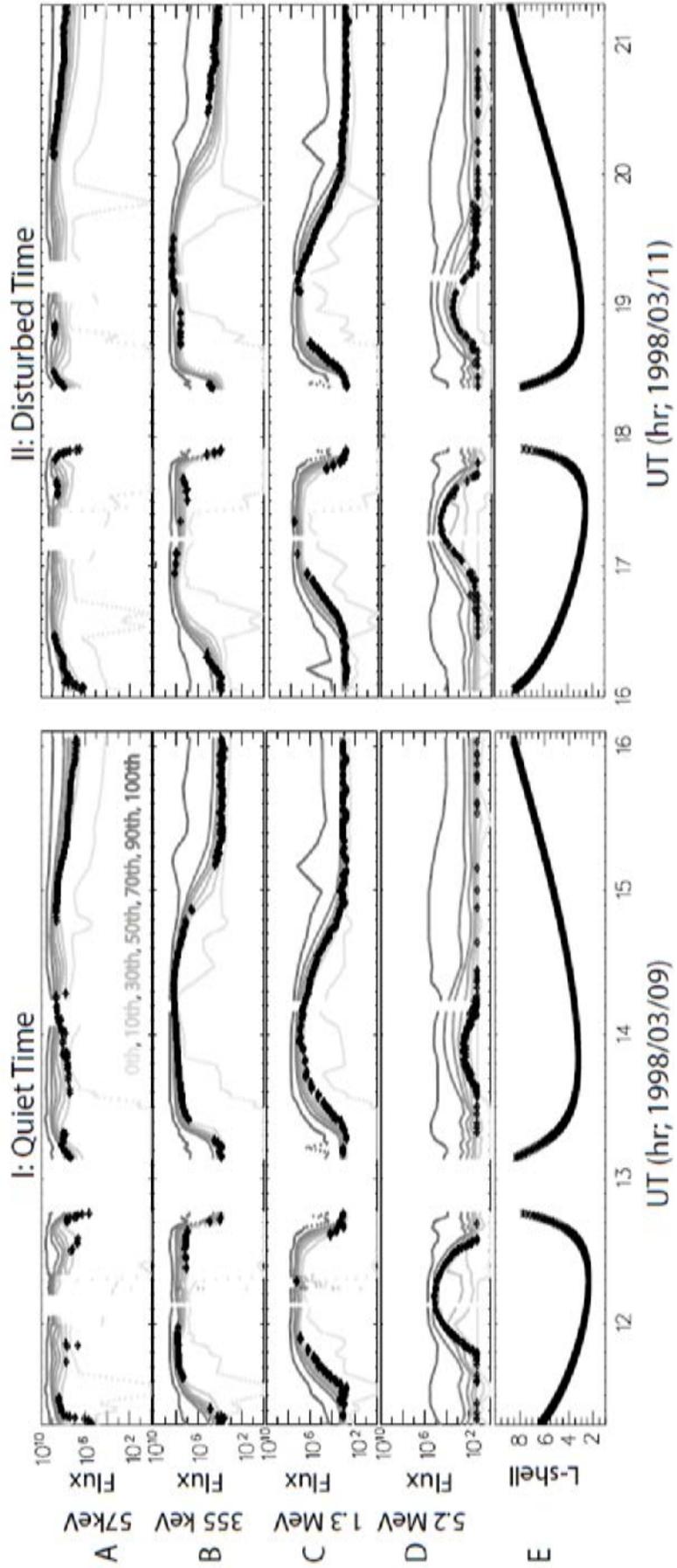

**Figure 7 Model validations: Comparing to Polar CEPPAD data. Column I)** Panels A to D show time series of proton fluxes (black diamond symbols, in /cm^2/s/sr/MeV) comparing to PolarP model output at four energy points: 57keV, 355keV, 1.3MeV, and 5.2MeV, respectively, during a ~5 hr period on a geomagnetically quiet day (1998/03/09): In each panel, PolarP fluxes (in gray) are given at seven percentiles: $0^{th}$ (i.e., the minimum for the bin), $10^{th}$, $30^{th}$, $50^{th}$, $70^{th}$, $90^{th}$, and $100^{th}$ (i.e., the maximum), respectively. Panel E shows L-shells during the period. **Column II)** Comparisons during a ~5 hr period on a geomagnetically disturbed day (1998/03/11).

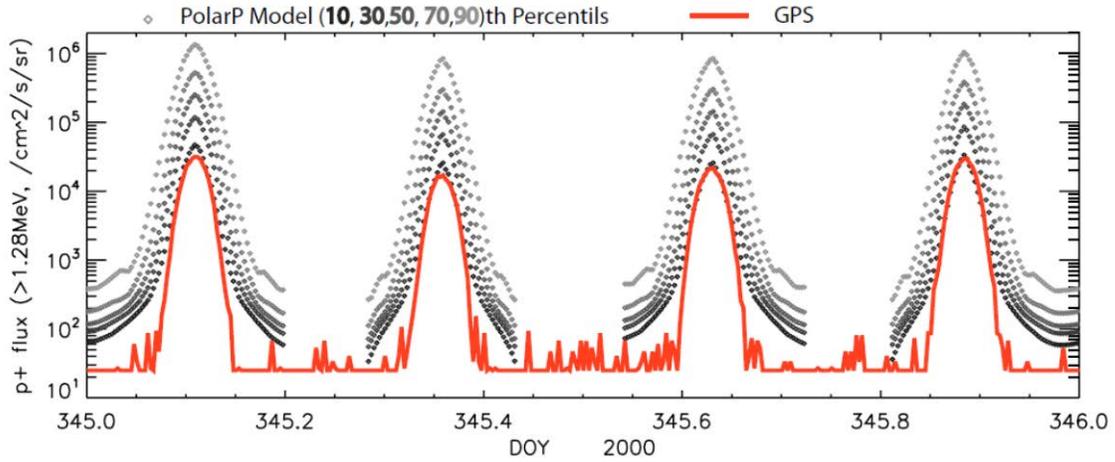

**Figure 8 Model validations: Comparison to GPS ns41 BDD-IIR data.** Time series data of GPS ns41 proton fluxes (red) with the energy within [1.27, 5.3] MeV compare to PolarP model fluxes (in gray) at five percentiles, $10^{th}$, $30^{th}$, $50^{th}$, $70^{th}$, and $90^{th}$, respectively, on one geomagnetically quiet day.

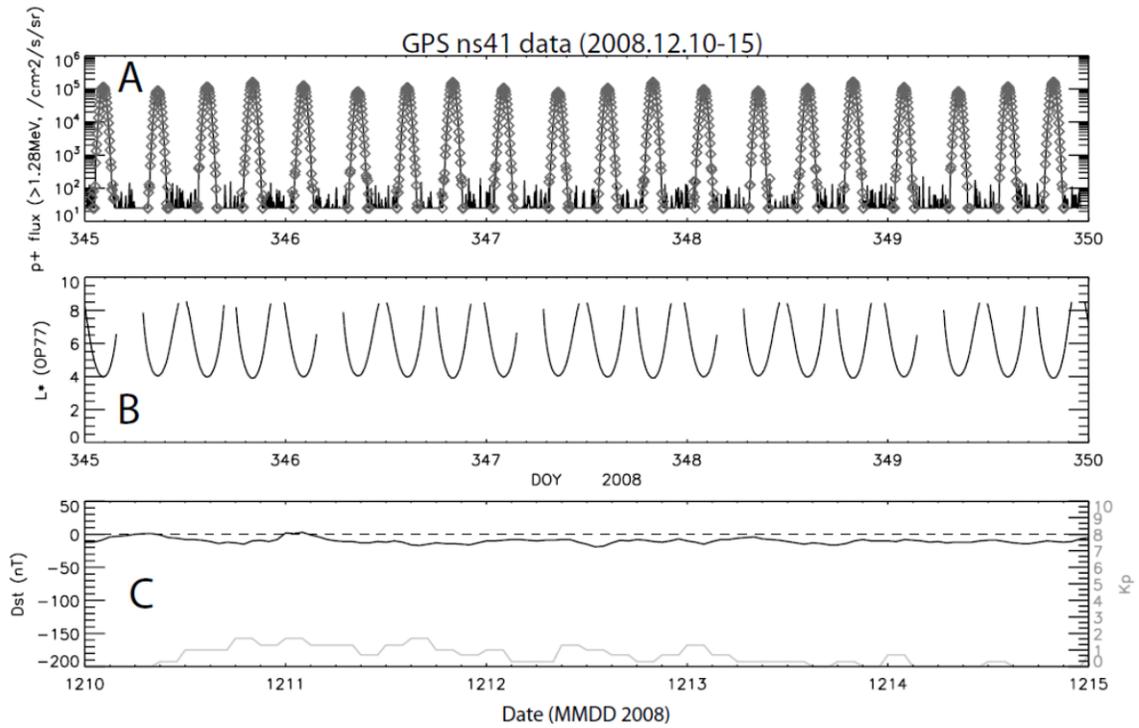

**Figure 9 A Sample of GPS ns41 BDD-IIR proton data with energy between [1.27, 5.3] MeV. A)** Time series ns41 proton fluxes (black curve) during a 5-day period (December 10 - 14, 2008). Data time resolution is 4 min. Data points measured with $L^* < 6$ are highlighted by diamond symbols in gray. **B)** $L^*$ calculated from the OP77 model. **C)** Dst (black) and Kp (gray) indices during the period.

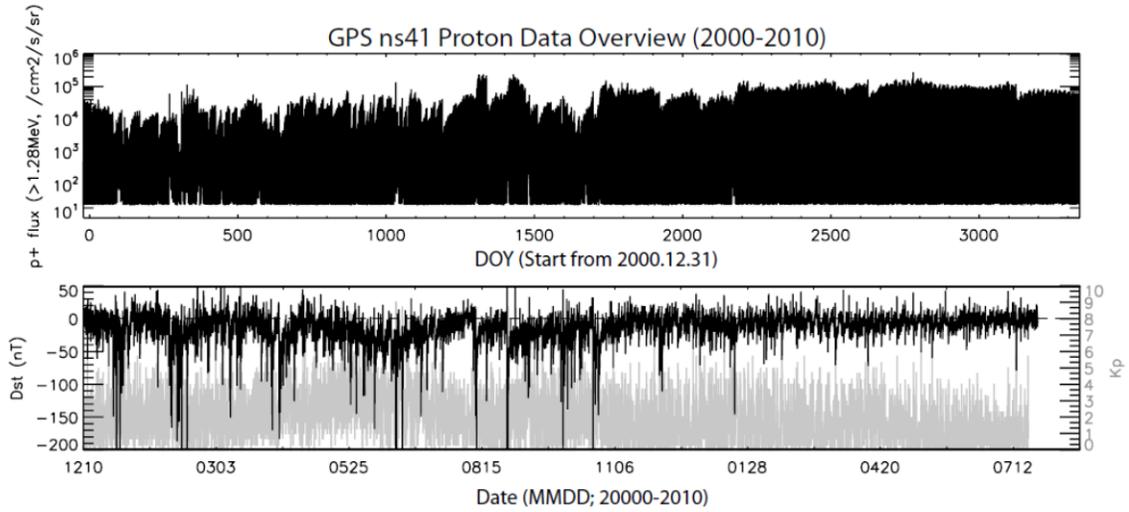

**Figure 10 Overview of GPS BDD-IIR proton data (2000-2010). A)** Time series of GPS ns41 proton fluxes (black curve) with energy between [1.27, 5.3] MeV from 2000 to present (2010). **B)** Dst (black) and Kp (gray) indices during the same period.

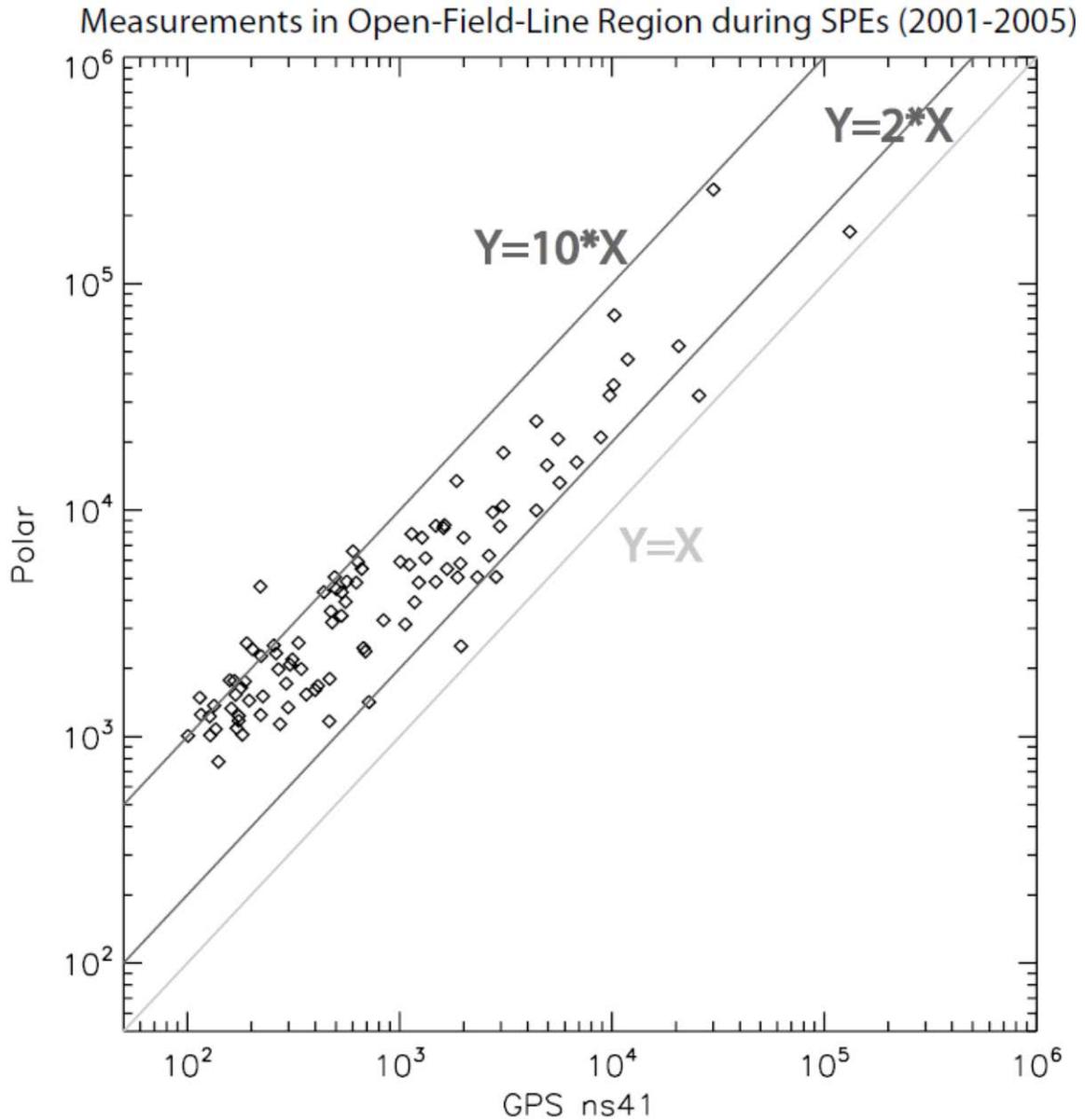

**Figure 11 Inter-instrument calibration between Polar CEPPAD and GPS ns41 BDD-IIR proton channel with energy between [1.27, 5.3] MeV.** Each data point (diamond symbols) represents the fluxes measured by both instruments in the open-field-region within a 2-hour time window during a Solar Proton Event. The flux unit is #/cm^2/s/sr. Three straight lines in gray: The one in diagonal indicates perfect inter-calibration between instruments; the other two indicate Polar fluxes are 2 and 10 times higher than GPS fluxes.

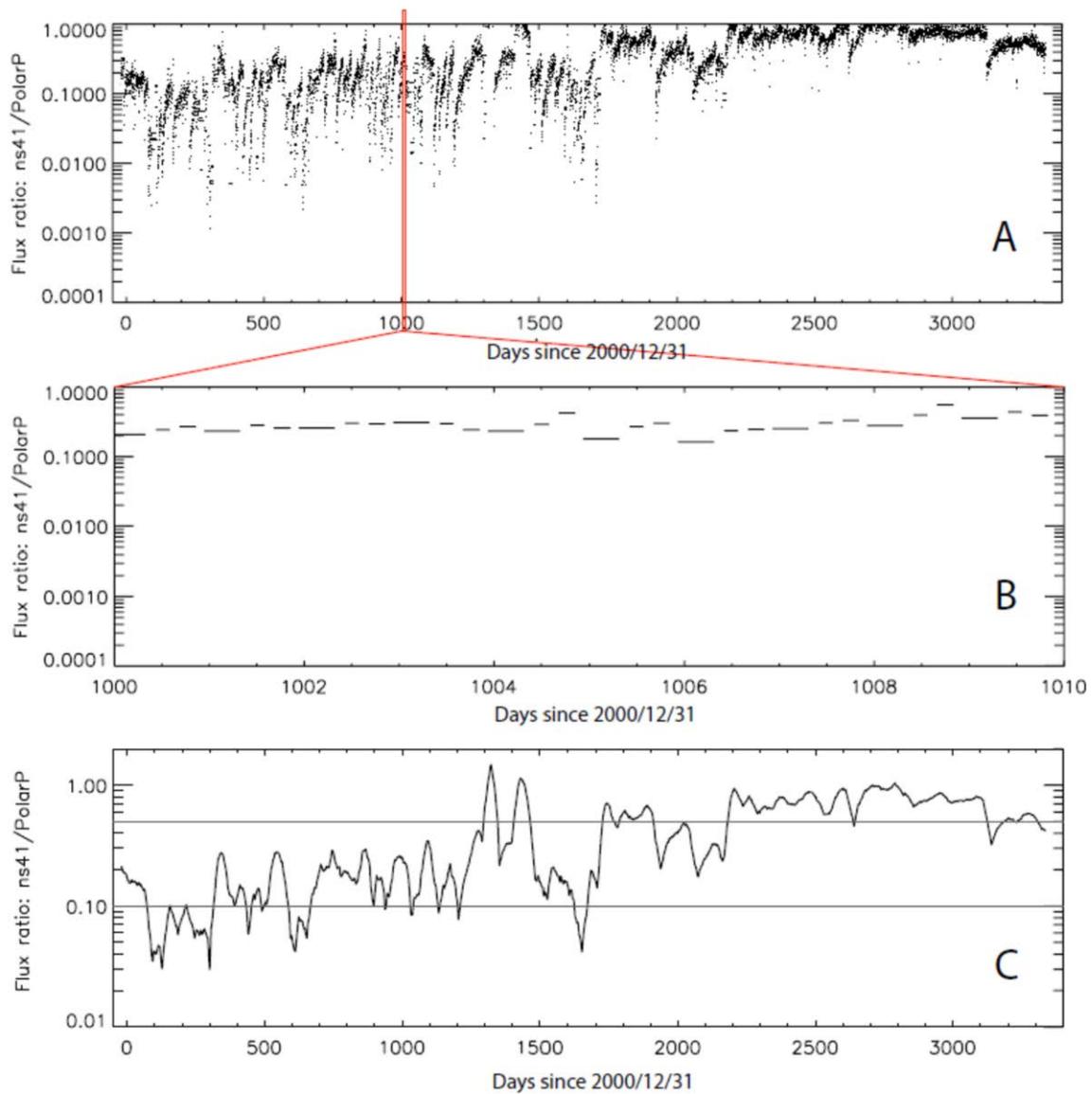

**Figure 12 Ratios between GPS ns41 BDD-IIR proton data and PolarP fluxes with energy between [1.27, 5.3] MeV. A)** Ratios for the whole period (2000 - 2010). A 10-day interval is boxed out (red) and enlarged in next panel to show details**. B)** Ratios during a 10-day interval. Each short straight line indicates the ratio during one (or two consecutive) radiation belt crossing(s) by the GPS ns41. **C)** Averaged ratios (black) running over 27-day window for the whole period (2000 - 2010). Two horizontal lines indicate ratio levels of 0.1 and 0.5. Note here the vertical scale is different from the other two panels.

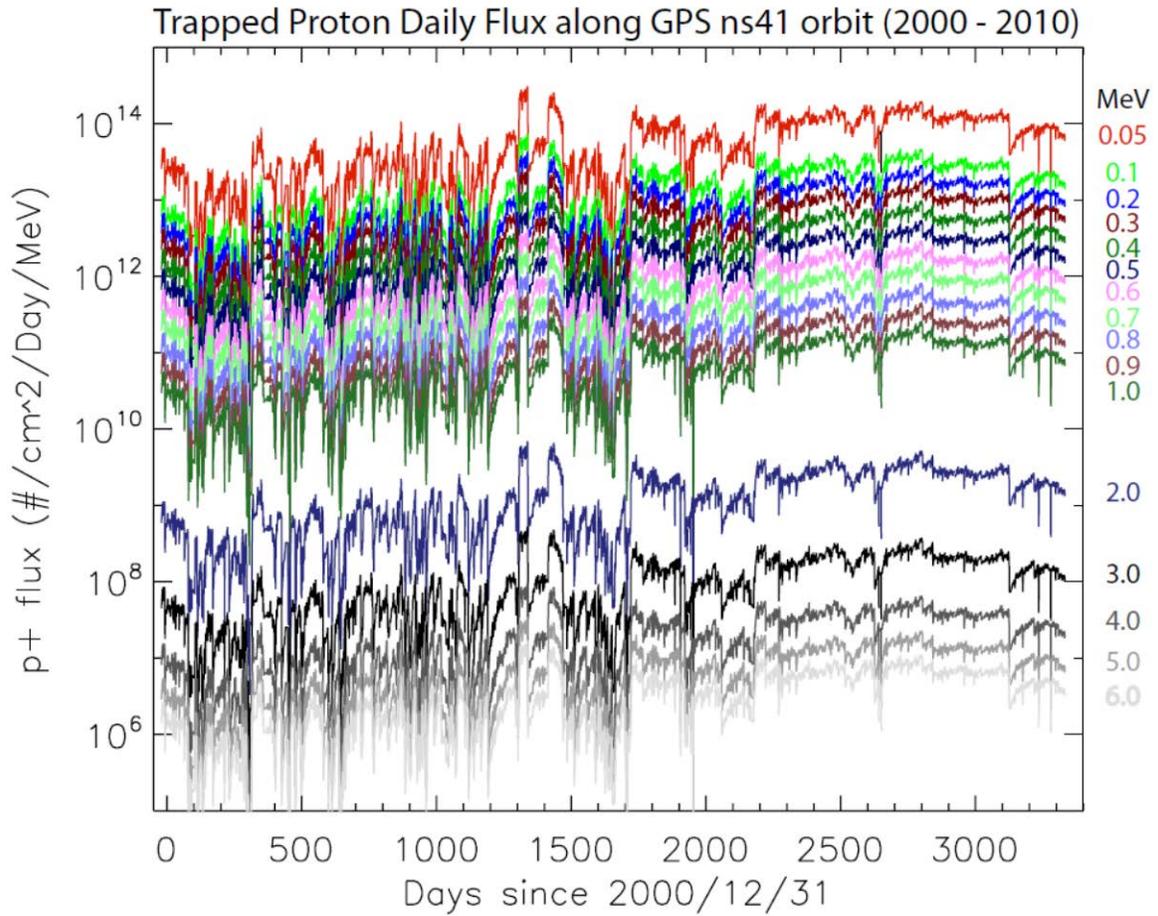

**Figure 13 Derived time series of trapped proton fluxes at 16 energy points along the GPS ns41 orbit (2000 – 2010).** The flux unit is /cm^2/s/Day/MeV. Values of the 16 energy points are presented to the right of the panel, with the same color as the corresponding curve. Fluxes presented here are derived from PolarP fluxes at the 50$^{th}$ percentile (see text for details).

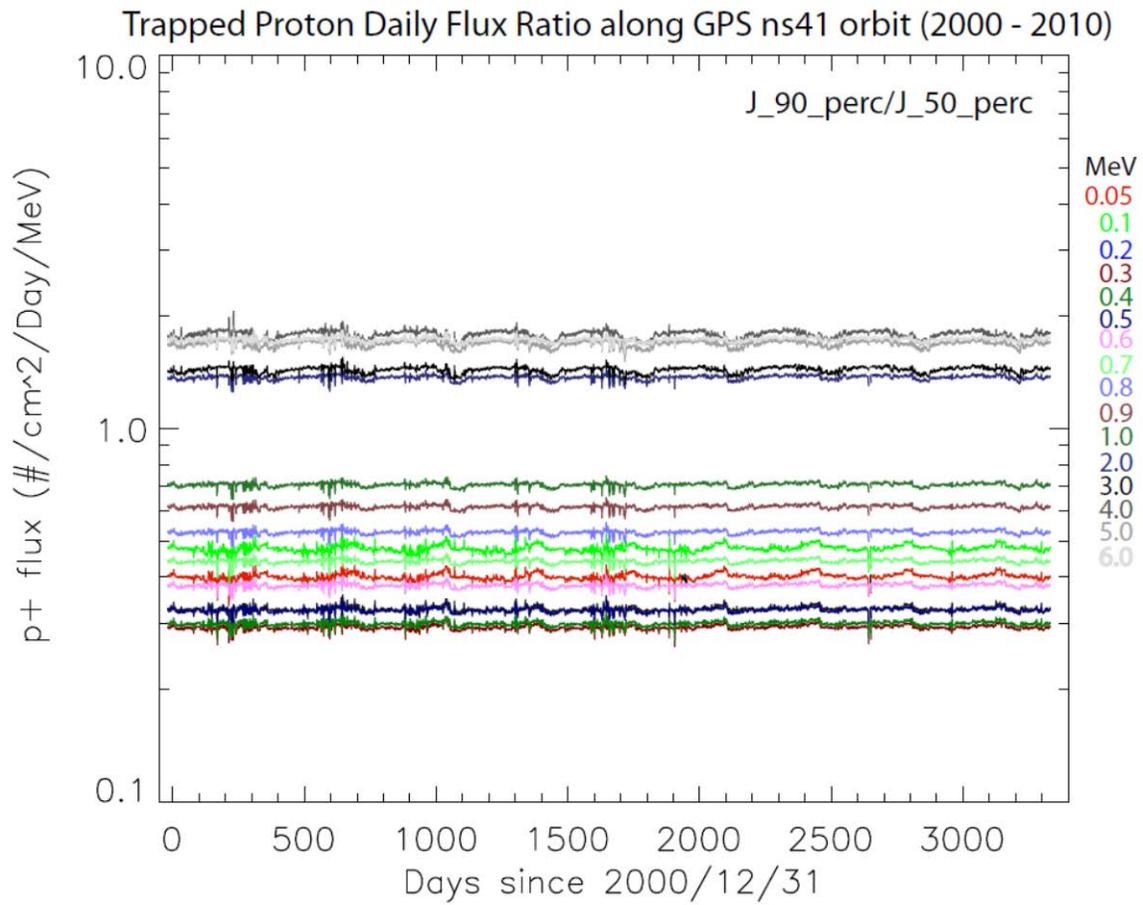

**Figure 14** Ratios between derived proton fluxes at 16 energy points along the GPS ns41 orbit based upon PolarP fluxes at the 90$^{th}$ and 50$^{th}$ percentiles (2000-2010).

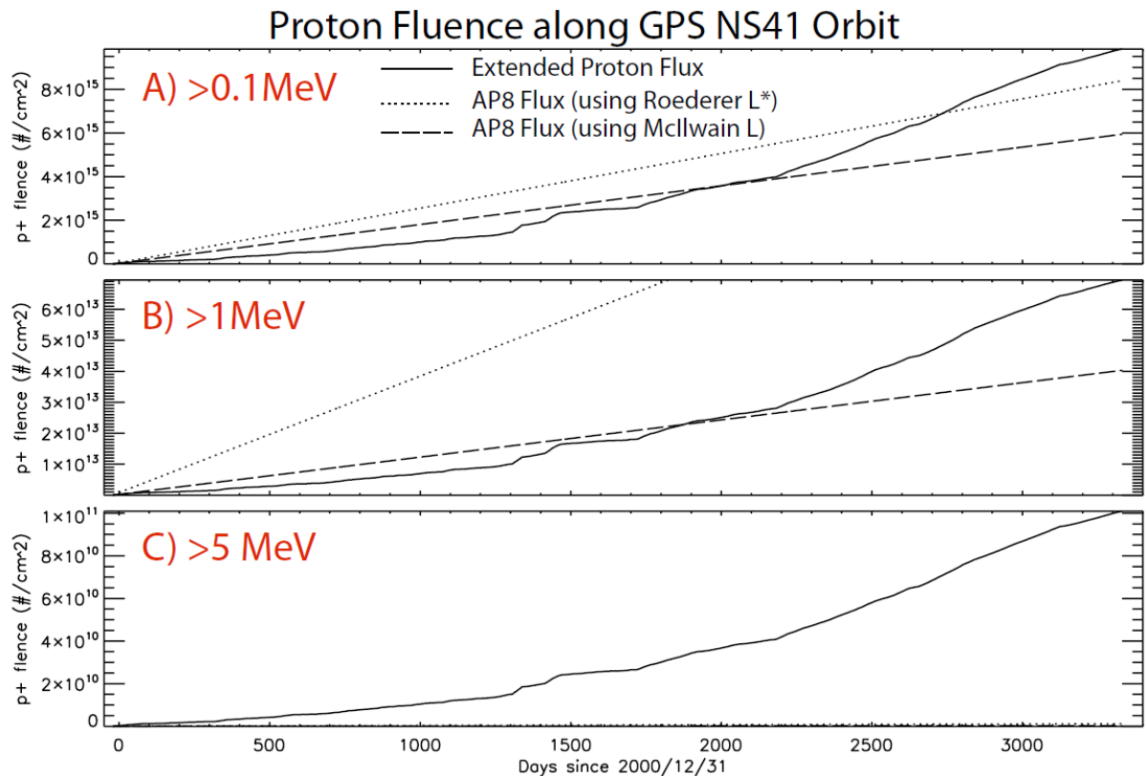

**Figure 15 Fluence accumulated from derived proton fluxes (solid lines) during the whole period (2000-2010) comparing to the results based on AP8 Min model (dotted and dashed lines). A)** Accumulated fluence for protons with energy above 0.1 MeV. The dotted line shows AP8 results by using Roederer L-shell values from the OP77 model as the input parameter, and the dashed line shows AP8 results by using McIlwain L parameters from the OP77 model as the input parameter. **B)** Above 1MeV, and **C)** above 5MeV. The AP8 curves in Panel C are so close to the horizontal axis that can be barely seen here.